\documentclass{PoS}

\newcommand{\pt}{\ensuremath{p_{\mathrm{T}}}}
\newcommand{\kt}{\ensuremath{k_{\mathrm{T}}}}
\newcommand{\MeV}{\,\ensuremath{\mathrm{Me\kern-0.1em V}}}
\newcommand{\GeV}{\,\ensuremath{\mathrm{Ge\kern-0.1em V}}}
\newcommand{\TeV}{\,\ensuremath{\mathrm{Te\kern-0.1em V}}}

\title{Jet Reconstruction with charged tracks only in CMS}

\ShortTitle{Jets finding with tracks in CMS}

\author{Paolo Azzurri \thanks{on behalf of the CMS Collaboration}\\
        Scuola Normale Superiore, piazza dei Cavalieri 7,  56100 Pisa, Italy \\
        E-mail: \email{paolo.azzurri@cern.ch}}

\abstract{The performance of jet finding using only charged tracks in CMS 
has been investigated. Different jet algorithms have been applied to QCD 
di-jet events, to hadronic $t\overline{t}$ multi-jet 
events and on Z+jets events.
Results using jets made with tracks only
or calorimeter towers are compared 
for energy response, angular resolution and
jet matching to the leading partons.  
The jet reconstruction performance in the presence of pile-up interactions 
is presented for the Z+jets sample.
}

\FullConference{2008 Physics at LHC\\
		 29 September - 4 October 2008\\
		 Split, Croatia}

\begin{document}

\section{Introduction\label{sec:intro}}

The performance of jet finding in CMS based on clustering the energy 
deposits of calorimeter towers (``CaloTowers'') into calorimeter jets (``CaloJets") 
using different jet algorithms and jet size parameters 
is well documented~\cite{ptdr,JME-07-003}. 
The studies are based on comparisons to jets clustered with the same algorithms 
with stable Monte~Carlo (MC) truth particles as inputs, which are referred to as generator jets ("GenJets").

Besides the fluctuations in the charged component, 
charged particles represent the jet component that is 
measured best both in terms of 
energy resolution and of angular direction,
that is extremely well determined at the interaction point.
For these reasons,
one can expect the tracks picture of a 
multi-jet event to be more collimated than the CaloTower picture, 
with less overlap and interference between jets, 
and less background (e.g. from pile-up events). 

With respect to CaloJets, jet reconstruction 
with charged tracks only (``TrackJets") is 
a completely independent 
way to find and count jets, and determine their directions,
with independent detector-related 
systematic uncertainties~\cite{JME-08-001}.

\vspace{-0.3cm}
\section{\label{sec:qcd} Performances in QCD di-jet events}

In order to study the performance of jet reconstruction with charged tracks only,
jets are clustered using
{\it a)} the standard Iterative Cone (IC) algorithm and 
{\it b)} the $\kt$ algorithm.
The ``E-scheme'' is used for the clustering recombination.
The Iterative Cone algorithm is used with different radii from 
$R=0.3$, to $R=0.7$. 
The $\kt$ jets are clustered with size-parameters from $D=0.2$ to $D=0.6$.
Reconstructed TrackJets are compared to CaloJets 
and GenJets using the same schemes.

\begin{figure}[hbtp]
  \begin{center}
      \includegraphics[width=0.48\textwidth]{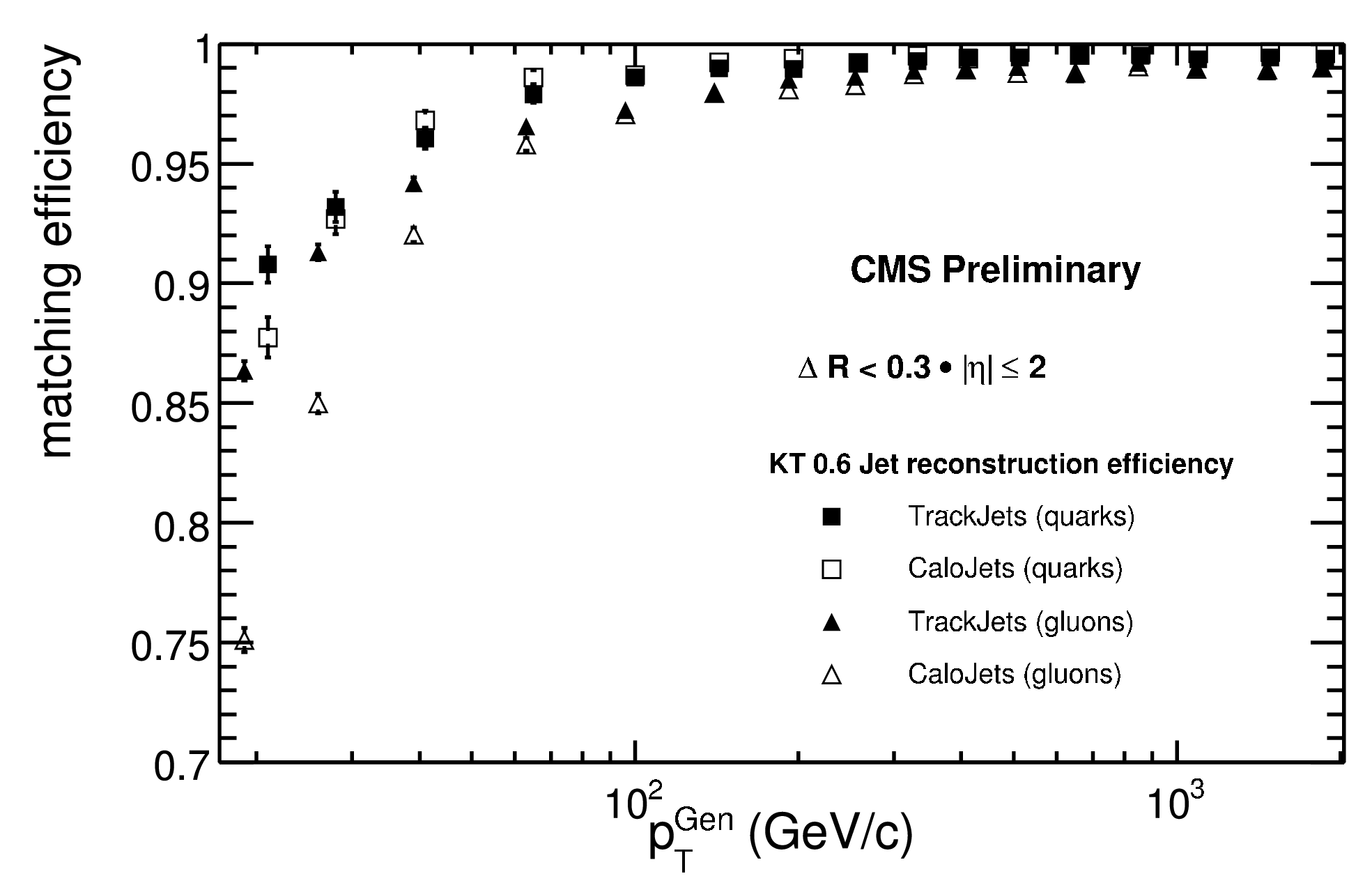}
            \includegraphics[width=0.48\textwidth]{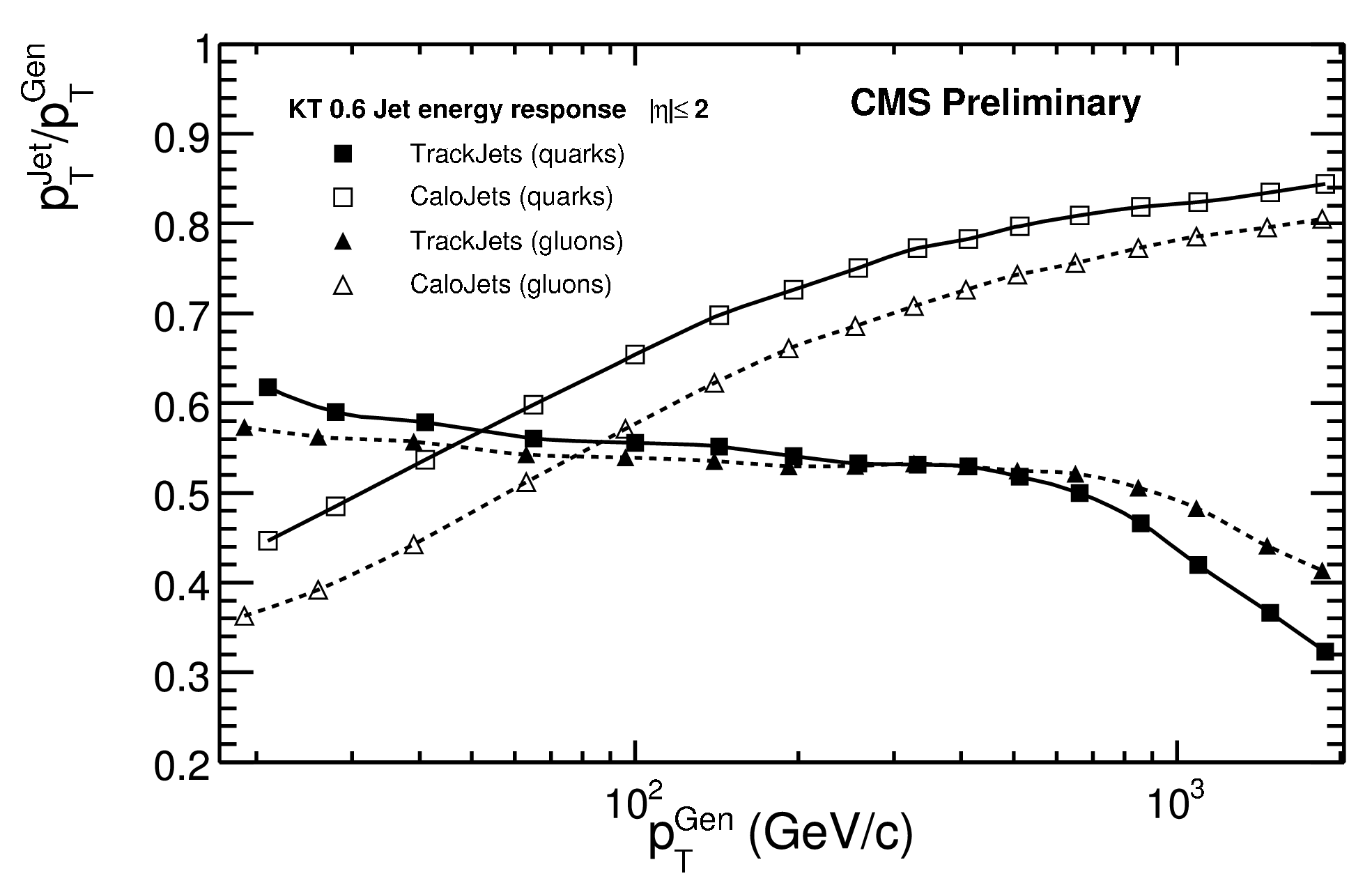}
    \caption{Reconstruction efficiency and $\pt$ response with respect to Monte~Carlo truth jets associated
to the leading parton in QCD di-jet events, shown separately for jets originating from quarks and gluons.
}
    \label{fig:jetresp}
  \end{center}
\end{figure}

Results are obtained studying QCD di-jet events where the leading $\pt$ parton  
is well within the tracker acceptance ($|\eta|<2$) 
and is associated to a GenJet. 
Figure~\ref{fig:jetresp} shows the efficiencies to find a reconstructed jet
within $\Delta R<0.3$ to the GenJet associated to the leading parton,
for CaloJets and TrackJets, using a $\kt$ clustering algorithm
with $D=0.6$.
TrackJets have a slightly better efficiency than CaloJets
at low $\pt$. 

\vspace{-0.5cm}

Figure~\ref{fig:jetresp} also shows that the CaloJet \pt \ response increases almost linearly as
a function of $\log \pt$ while the TrackJet response is rather stable around 
55\%, decreasing slowly in the $\pt$=100 \GeV/c to  1~$\TeV$/c range 
and dropping for $\pt>$1~$\TeV$/c, due to tracking inefficiencies
in the core of high $\pt$ jets.
It also can be seen that the response to gluon jets is lower than for quark jets,
because of their wider shower and softer fragmentation function.


The angular resolution of the reconstructed jet axis is 
evaluated with respect to the direction of the 
Monte~Carlo truth GenJet. 
TrackJets yield better $\phi$ resolutions for $\pt <200\GeV$/c,
while the CaloJets $\phi$ resolutions improve at higher \pt.
The $\eta$ resolutions of TrackJets and CaloJets are very similar,
where the $\eta$ direction of the CaloJets is corrected to the position
of the primary interaction vertex determined using the 
reconstructed tracks.

\begin{figure}[hbtp]
  \begin{center}
      \includegraphics[width=0.48\textwidth]{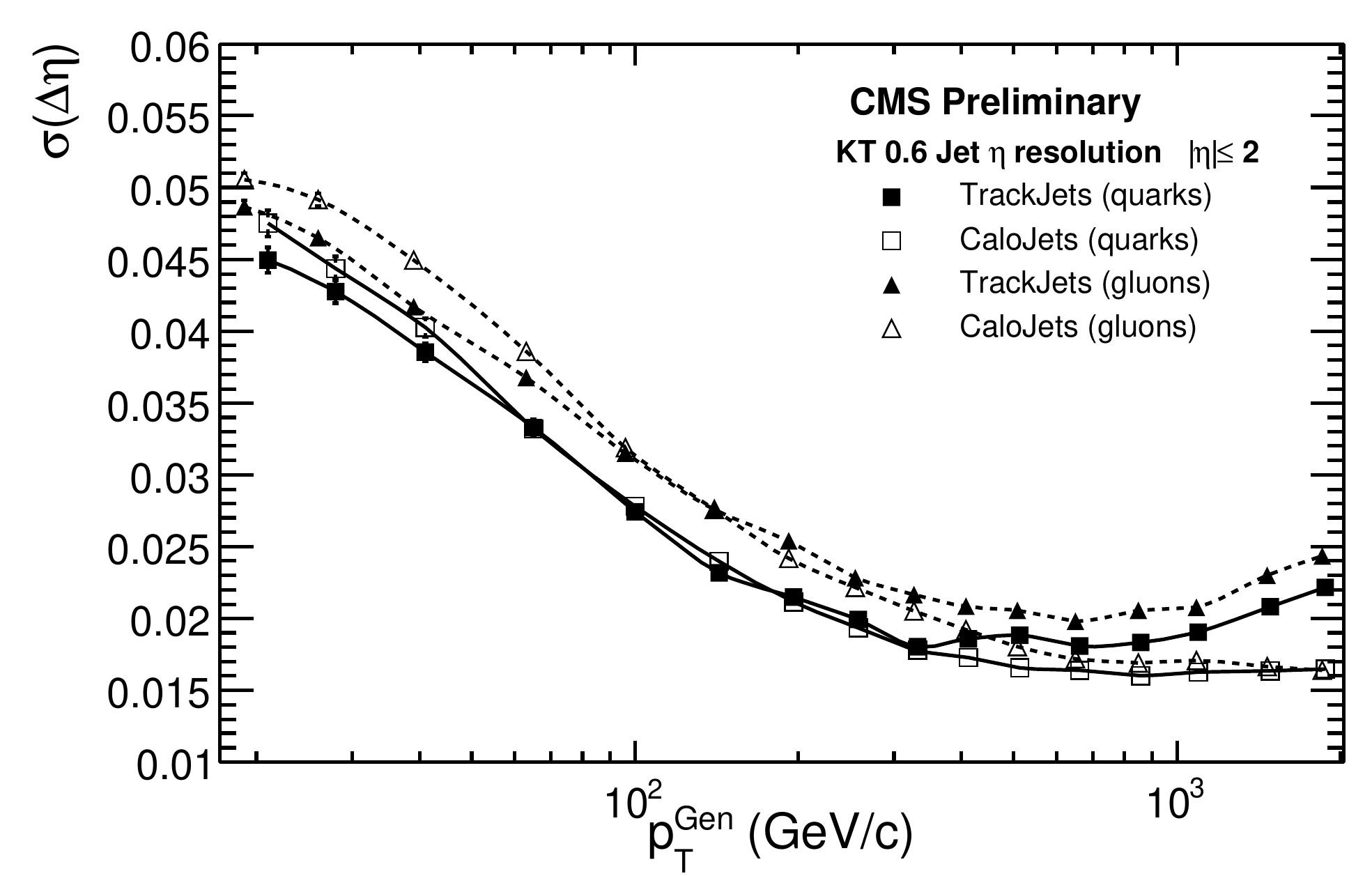}
            \includegraphics[width=0.48\textwidth]{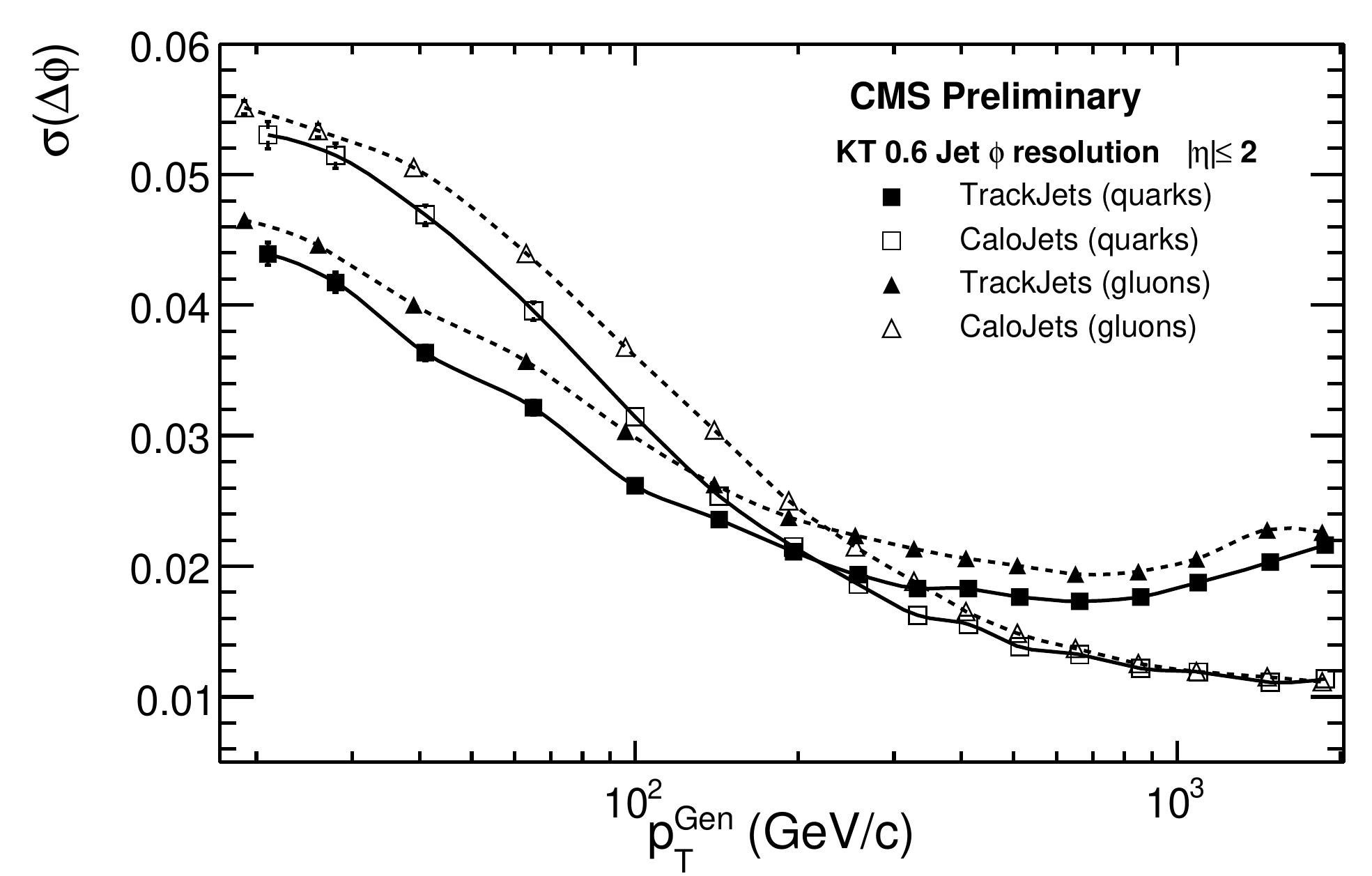}
    \caption{Jet $\eta$ and $\phi$ resolution as a function of the 
      GenJet $\pt$, for calorimeter and TrackJets 
      originating from quarks and gluons.
      All jets are clustered with the $\kt$ algorithm with $D=0.6$.
      }
    \label{fig:jetang}
  \end{center}
\end{figure}
\vspace{-0.5cm}
\section{\label{sec:tt} Performances in hadronic top pairs decays}
To evaluate the performance of jet clustering with tracks only in a multi-jet
environment, samples of fully hadronic decays of $t\overline{t}$ events
are analyzed.
The final state of these events consists of six quarks 
($b\overline{b}q\overline{q}'q\overline{q}'$)
that should be reconstructed as six jets.

The matching of jets to quarks is done for each reconstructed jet 
collection associating each quark to the closest jet in 
$\Delta R$ and requiring 
$\Delta R<$0.2.
Quarks that match to the same jet indicate that their jets have been 
merged, so they are removed from the 
list of matched quarks.

The optimal jet clustering size turns out to be  
in the $R=0.3-0.5$ range for the cone jets radius
and in the $D=0.4-0.6$ range for the $\kt$ jets clustering, 
and the best matching 
performances are obtained with TrackJets.

\section{\label{sec:zj} Performances in Z+jets events}
Direct productions of vector bosons have large cross sections at the LHC, and 
multi-jet events associated with 
W/Z bosons are relevant to Standard Model (SM) measurements and for 
searches beyond the SM.   
In these events TrackJets can provide a robust 
method for jet counting.

Events where the Z boson decays into a pair of muons with both muons 
reconstructed with $p_{\rm T}(\mu) > 20 ~{\rm \GeV}$/c have been studied.
The jet finding efficiency turns out to be significantly better for TrackJets 
for $\pt<30\,\GeV$/c and comparable to CaloJets for higher 
GenJet $\pt$.

Another useful measure of the quality of jet reconstruction is the 
jet mismatched rate, here defined as the 
fraction of the reconstructed jets, which are not matched to a 
generator level jet within $\Delta R <0.3$.

\begin{figure}[hbtp]
  \begin{center}
      \includegraphics[width=0.48\textwidth]{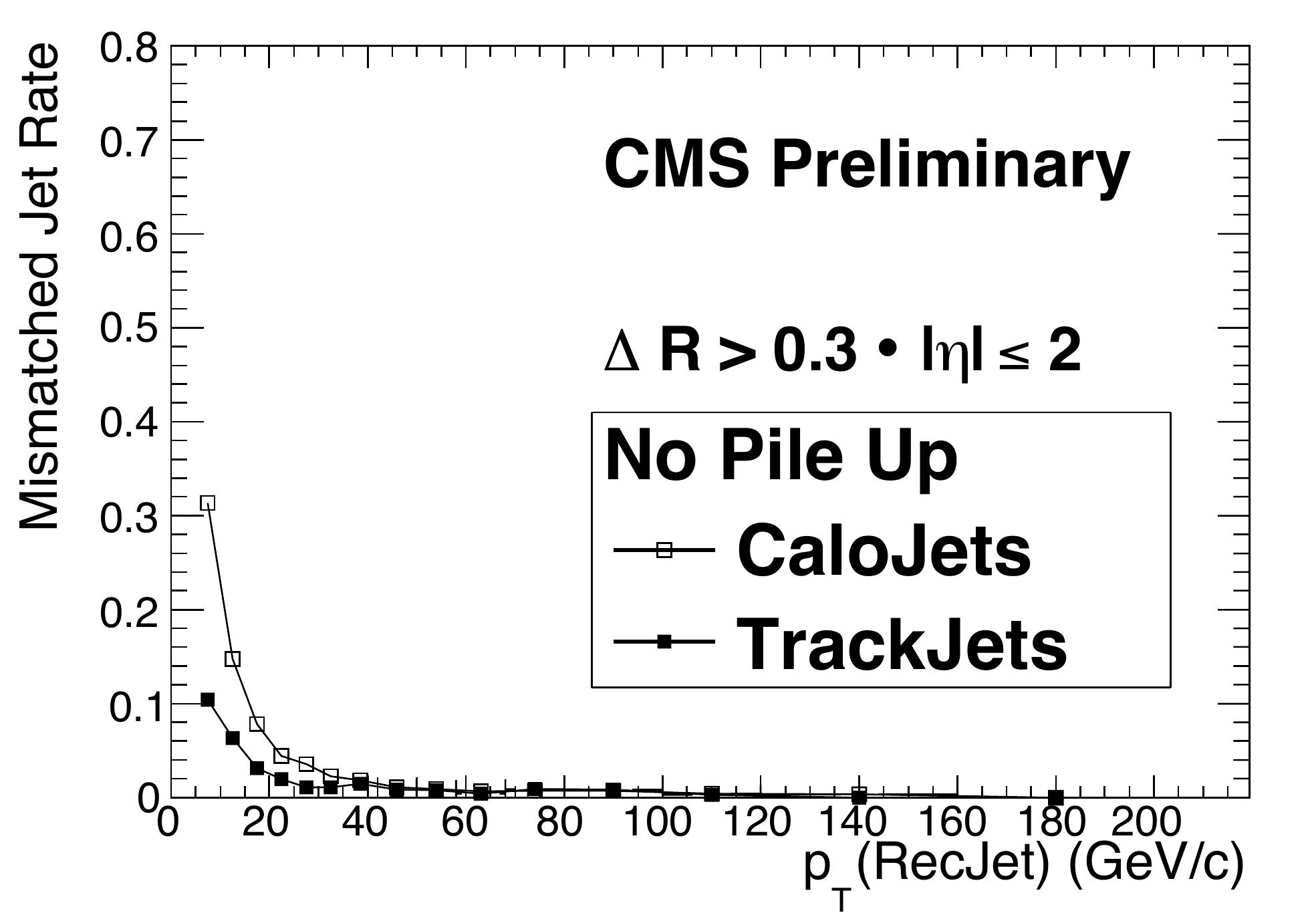}
      \includegraphics[width=0.48\textwidth]{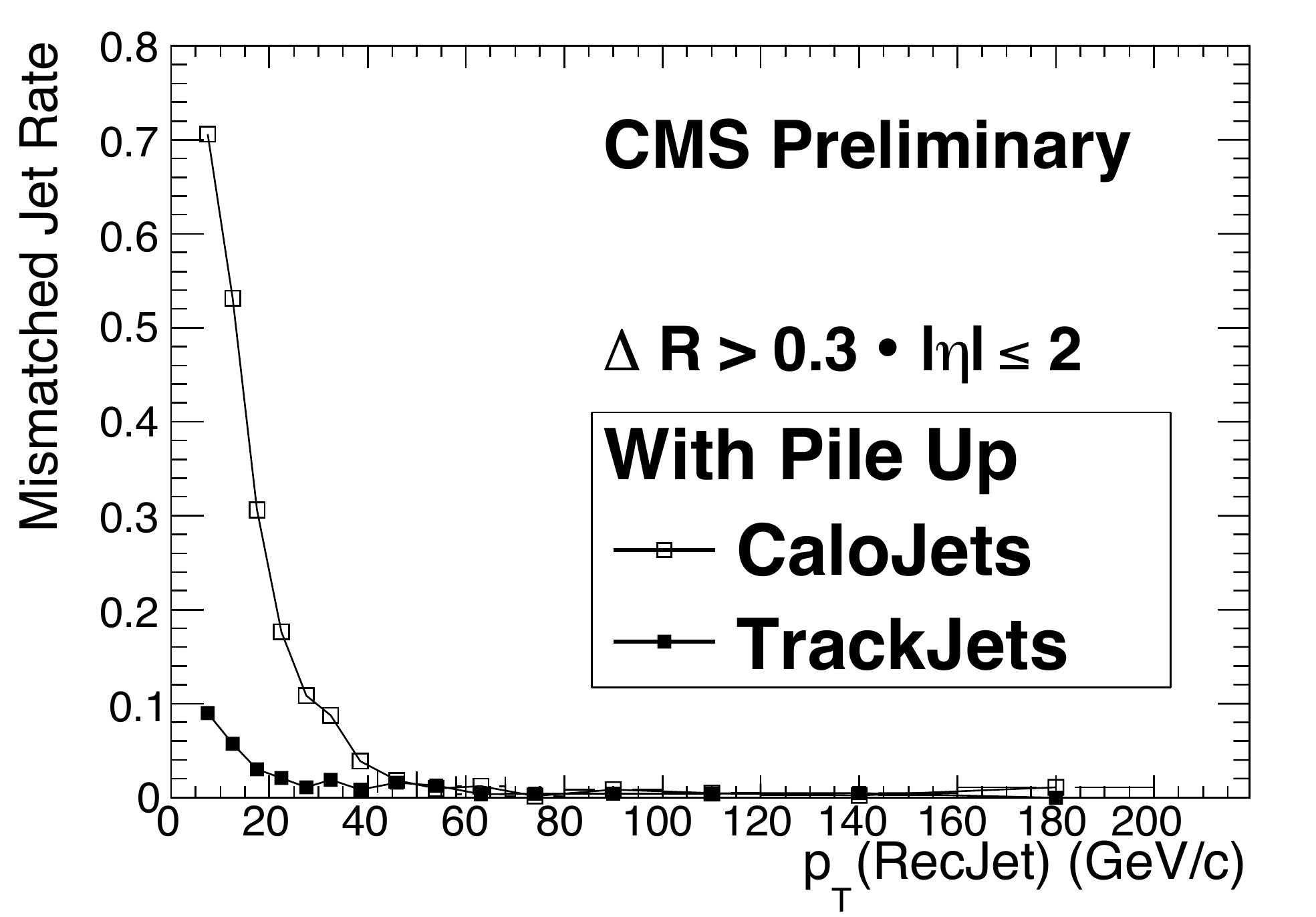}
       \caption{Jet mismatched rate in Z+jets events, as a function of the reconstructed jet $\pt$,
      where signal events are without or with pile-up events.
      }
    \label{fig:zjets}
  \end{center}
\end{figure}

Figure~\ref{fig:zjets} shows the mismatched jet 
rate for the TrackJets and 
CaloJets. 
The TrackJets give smaller mismatched jet rate 
(less than a few $\%$ for $\pt> 15~{\rm \GeV}$/c) 
than the CaloJets, and
the difference is particularly pronounced in
the low jet $\pt$ region. 

\vspace{-0.5cm}

To study the effect of pile-up, 
the signal samples of Z+jets are overlayed 
with a sample of Minimum Bias events where 
the mixing proportion is 1:5 per bunch
crossing on average.

Tracks are measured at the interaction point and the 
information about the vertex of origin is therefore naturally taken into account.
This information can be used
to reject tracks that are not compatible with the 
primary interaction point, in this case the di-muon vertex.

Figure~\ref{fig:zjets}  also
shows the jet mismatched rate after the pile-up events are included and
reveals that the mismatched rate for TrackJets is not affected by 
the addition of pile-up events, while it increases significantly for CaloJets. 

\section{Conclusions\label{sec:conc}}
The performance of jet reconstruction with charged tracks only in CMS (TrackJets) has 
been presented and compared to calorimeter jets (CaloJets).
Within the tracker acceptance,
the jet finding efficiency with TrackJets is higher
than for CaloJets for $\pt\leq 30$~$\GeV$/c, and the angular resolution
of TrackJets is better than for CaloJets 
at $\pt\leq 200$~$\GeV$/c, mainly for the azimuthal $\phi$ resolution.

In multi-jet events, such as $t\overline{t}$
decays to six jets,
the performance of jet finding with
TrackJets is better than with CaloJets
when considering the efficiency to match correctly the 
six quarks from the top decays, 
while setting the mismatched rate to the same level.

In Z+jets events 
TrackJets provide a 
simple and clean way for jet  counting. 
Using the track vertex information 
the reconstruction of TrackJets is not affected by pile-up events.

\end{document}